\documentclass[twocolumn,superscriptaddress,secnumarabic,amssymb, nobibnotes, aps, prb]{revtex4-2}

\usepackage{amsmath}    
\usepackage{graphicx}    
\usepackage{verbatim}    
\usepackage{color}           
\usepackage{subfigure}   
\usepackage{hyperref}    
\usepackage{braket}
\usepackage{physics}
\usepackage{xfrac}
\usepackage{xspace}
\newcommand{\amt}{$\alpha$-MnTe\xspace}
\newcommand{\ruo}{RuO$_2$\xspace}
\newcommand{\mf}{MnF$_2$\xspace}
\newcommand{\bk}{\mathbf{k}}
\newcommand{\hbk}{\hat{\mathbf{k}}}

\newcommand{\bS}{\mathbf{S}}
\newcommand{\bh}{\mathbf{h}}
\newcommand{\bbm}{\mathbf{m}}

\newcommand{\bL}{\mathbf{L}}
\newcommand{\hbL}{\hat{\mathbf{L}}}
\newcommand{\na}{[110]\xspace}

\newcommand{\nc}{[001]\xspace}
\def\bL{\mathbf{L}}

\def\hT{\hat{T}}

\begin{document}
\title{
Determination of the N\'eel vector in rutile altermagnets through x-ray magnetic circular dichroism: the case of \mf}

\author{A.~Hariki}
\affiliation{Department of Physics and Electronics, Graduate School of Engineering,
Osaka Metropolitan University, 1-1 Gakuen-cho, Nakaku, Sakai, Osaka 599-8531, Japan}
\author{T.~Okauchi}
\affiliation{Department of Physics and Electronics, Graduate School of Engineering,
Osaka Metropolitan University, 1-1 Gakuen-cho, Nakaku, Sakai, Osaka 599-8531, Japan}
\author{Y.~Takahashi}
\affiliation{Department of Physics and Electronics, Graduate School of Engineering,
Osaka Metropolitan University, 1-1 Gakuen-cho, Nakaku, Sakai, Osaka 599-8531, Japan}
\author{J.~Kune\v{s}}
\affiliation{Department of Condensed Matter Physics, Faculty of
  Science, Masaryk University, Kotl\'a\v{r}sk\'a 2, 611 37 Brno,
  Czechia}

\begin{abstract}
We present a numerical simulation of x-ray magnetic circular dichroism (XMCD) at the $L_{2,3}$ edge of Mn in a representative 
rutile altermagnet \mf using
a combination of density functional theory + exact 
diagonalization of the atomic model. 
We explore how the dichroic spectra vary with the orientation of the light propagation vector and the N\'eel vector. 
An exact relationship between the XMCD spectra for different
orientations of the N\'eel vector, valid in the absence of
the valence spin-orbit coupling and core-valence multipole interaction,
is derived and its approximate validity for the full Hamiltonian verified by numerical calculation.
This relationship allows to determine the in-plane orientation of the N\'eel vector
using the XMCD spectra alone.

\end{abstract}

\maketitle


{\it Introduction.}~Altermagnets, a new species on the magnetism evolutionary tree,
have split from antiferromagnets recently~\cite{Smejkal22a, Smejkal22}.
Owing to their space-group symmetry altermagnets facilitate the presence of spin-polarized bands~\cite{Ahn19,Hayami19,Smejkal22a,Smejkal20,Yuan20,Yuan21,Hayami20,Smejkal22,Mazin21,Liu22,Jian23}, anomalous Hall effect~\cite{Smejkal22b,Smejkal20,Samanta20,Naka20,Hayami21,Mazin21,Gonzalez2023,Naka22},
odd magneto-optical effects~\cite{Naka20,Hariki2024a,Hariki2024b} and 
number of other phenomena~\cite{Watanabe2024} with odd N\'eel vector dependence.
The key element is the rotational or mirror symmetry, which links atoms on the distinct magnetic sublattices. This is different from the translational or inversion symmetry seen in conventional antiferromagnets and results in time-reversed states with opposite N\'eel vectors being macroscopically distinct.
The orientation of the N\'eel vector, including its sign, is thus an important question. Here, we show that the x-ray circular dichroism (XMCD) alone can answer it in some structures.

Much of the early studies on altermagnetism focused on RuO$_2$~\cite{Ahn19,Feng2022,Lovesey2022,Bai2022,Guo2024}, a metal with the rutile structure. Despite a considerable theoretical and experimental effort the altermagetism of RuO$_2$ is far from understood as the magnetic order in a bulk RuO$_2$ remains controversial~\cite{Berlijn17,Zhu2019,Smolyanyuk2024,Hiraishi24}. Moreover a large
magnetic field is needed to tilt the magnetic moments away from the $[001]$ easy-axis direction in order to allow finite odd magneto-optical effects such as circular dichroism~\cite{Hariki2024a,Sasabe23} or the anomalous Hall effect~\cite{Feng2022}.

These effects are facilitated by the spin-orbit coupling (SOC), which allows
the spin long-range order of non-relativistic origin to
influence the current response, observed in transport and optical experiments.
However, SOC acts also in the reverse direction. The coupling of spins to the current breaks the spin SU(2) symmetry, leading to the magneto-crystalline anisotropy or possibly inducing weak ferromagnetism as a result of canting of the local moments~\cite{Kluczyk2023}.
The latter disturbs the fully compensated magnetic state of an altermagnet  
and must be taken into account in the analysis of the experimental data~\cite{Feng2022}. 
XMCD 
takes advantage of SOC being naturally separated into a dominant core SOC and a minor valence SOC. The latter may give rise to a weak ferromagnetic XMCD signal, but its contribution to the XMCD spectra in lighter elements such as $3d$ metals is marginal~\cite{Hariki2024b,Kunes2003}.

Given the aforementioned uncertainty concerning the magnetism of \ruo~\cite{Smolyanyuk2024} it is desirable to establish the behavior of XMCD in a isostructural material that possesses a well-established magnetic order.
\mf is a perfect candidate~\cite{Yuan20}. It crystallizes in the rutile structure and its magnetism is thoroughly studied. The antiferromagnetic order sets in at around 67~K~\cite{Stout1942} with magnetic moments along the $[001]$ direction~\cite{Erickson1953,Keffer1952,Felcher96} similar to \ruo . 
In contrast to the metallic \ruo, \mf is a Mott insulator where the Mn$^{2+}$ configuration gives rise to a large spin moment of $S \approx 5/2$ and an orbital singlet.  Together with a small SOC in the $3d$ shell this results in a weak single-ion anisotropy.
When a strong enough magnetic field is applied along the $[001]$ direction, the moments reorient themselves perpendicularly to it, that is, into the $ab$-plane. Meanwhile, the sublattice magnetizations stay antiparallel to each other with a slight tilt towards the field direction. The field required for this spin-flop transition is around 9~T~\cite{Felcher96,King1979}. The direction of the N\'eel vector within the $ab$-plane is unknown.




{\it Methods.}~We perform a density functional theory (DFT) calculation for the experimental structure of MnF$_2$~\cite{Stout54} using the Wien2k package~\cite{wien2k}. The crystal field within the Mn 3$d$ shell is derived from the DFT bands using the Wannier90 and wien2wannier packages~\cite{wannier90,wien2wannier}, see the Supplemental Material (SM) for the computational details~\cite{sm}. 
Since MnF$_2$ is a large-gap Mott insulator, the Mn$^{2+}$ atomic model adequately accounts for the Mn $L_{2,3}$-edge XAS spectrum dominated by the intra-atomic multiplet effects as shown by the early studies by de Groot {\it et al.}~\cite{Groot90}. 
Recent experimental and theoretical work on isoelectronic MnTe~\cite{Hariki2024b}, which employed both the atomic model as well as 
the dynamical mean-field theory, came to the same conclusion. Therefore we use the atomic model where the lattice information is encoded 
in the sublattice-dependent crystal field. The staggered spin polarization described by the  N\'eel vector $\bL=\bS_1-\bS_2$ is generated 
by adding a Zeeman field of 0.01~eV in the desired direction, which is sufficient to achieve the saturated moment of approximately $5~\mu_{B}$.
We use a Hund's coupling constant $J = 0.86$~eV, which is standard for Mn$^{2+}$ systems~\cite{Hariki2024b,Anisimov91}. We include SOC in the Mn 2$p$ and 3$d$ shells and incorporate the Slater integrals for the 2$p$--3$d$ core-valence interaction, following the atomic Hartree-Fock calculation as described in Refs.~\onlinecite{Hariki2017,Hariki20}. More details can be found in the Supplemental Material (SM)~\cite{sm}.

\begin{figure}[t]
\includegraphics[width=0.93\columnwidth]{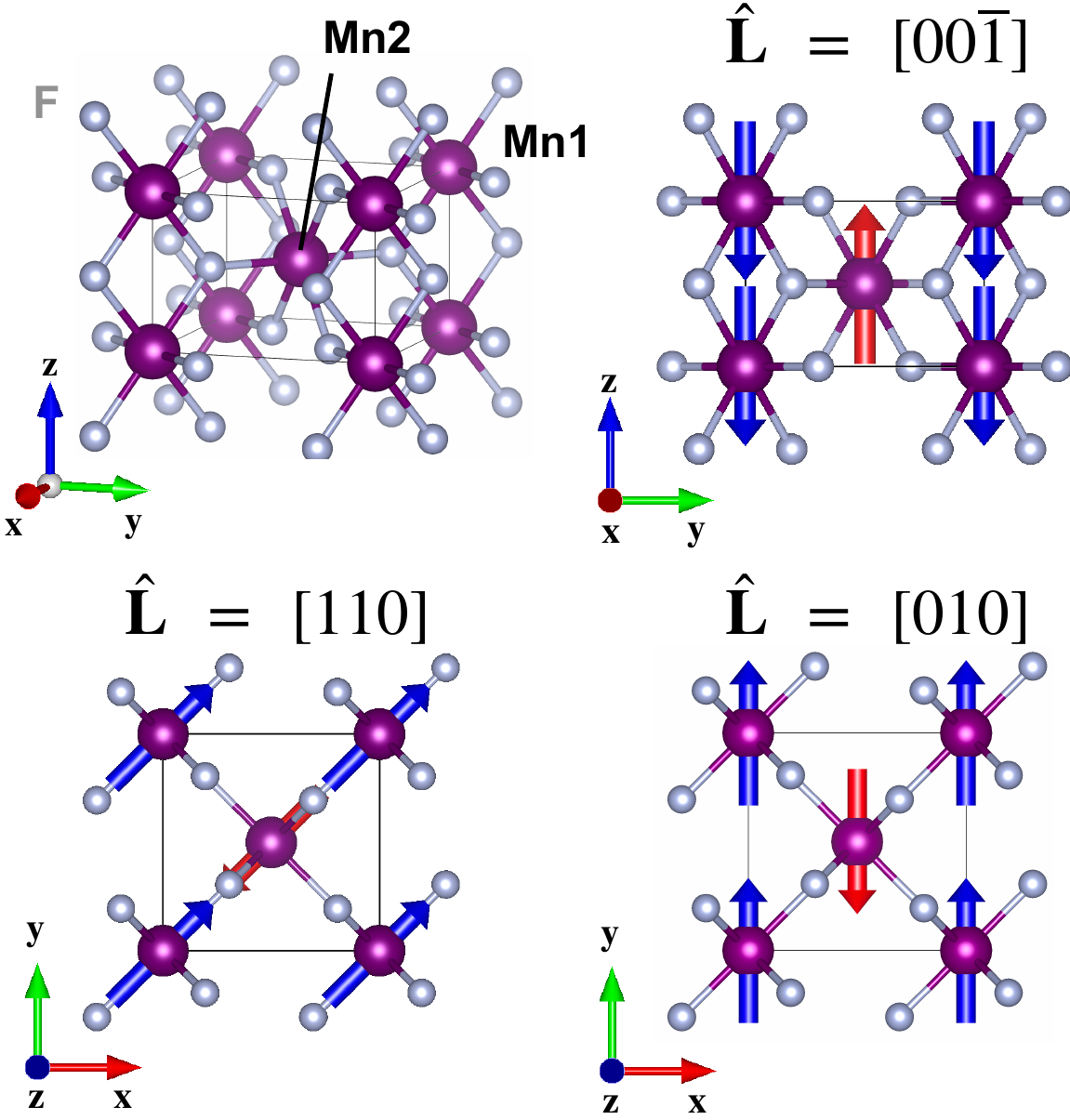}
\caption{
The rutile structure of \mf with various orientations of local spin moments.}
\label{fig_struct}
\end{figure}

The XMCD spectrum $\Delta F(\omega)=F^+(\omega)-F^-(\omega)$ is the difference of the absorption spectra for the right and left circularly polarized light 
propagating along the direction $\hat{\bk}$, 
obtained by the Fermi golden rule
\begin{equation}
\label{eq:fsum}
F^\pm(\omega;\hat{\bk},\bL)=\sum_{f} \left|\expval{f_\bL|\hT^{\pm}_{\hat{\bk}}|i_\bL}\right|^2\!\delta\left(\omega\!-\!E_{fi;\bL}\right).
\end{equation}
Here $|i_\bL\rangle$ and $|f_\bL\rangle$ are the eigenstates of the Hamiltonian, $E_{fi;\bL}$ is the excitation energy~\footnote{The spectra are calculated relative to the $L_3$ edge, the absolute excitation energies are obtained by matching the XAS spectra with experiment.}, and $\hT^{\pm}_{\hat{\bk}}$ are the dipole operators for the right- and left-hand polarization
with respect to the propagation vector $\hat{\bk}$~\footnote{In the following we use the hat symbol $\hat{\mathbf{v}}$ to indicate the direction of vector $\mathbf{v}$, i.e., $\mathbf{v}=a\hat{\mathbf{v}}$ for some $a>0$.}. Thanks to the immobility of the core hole
the x-ray absorption spectrum is a sum over the site contributions.
In the dipole approximation the dependence on $\bL$ and $\hat{\bk}$ has the form
${\Delta F(\omega;\hat{\bk},\bL)=2\operatorname{Im}\bh_\bL(\omega)\cdot \hat{\bk}}$. The axial vector ${\bh(\omega) = (\sigma^a_{zy}(\omega), \sigma^a_{xz}(\omega), \sigma^a_{yx}(\omega))}$, representing the antisymmetric part of the conductivity tensor $\boldsymbol{\sigma}(\omega)$, is the finite frequency equivalent of the Hall vector~\cite{Wimmer19, Smejkal22a, Hariki2024a}.


\begin{figure}
\includegraphics[width=1.00\columnwidth]{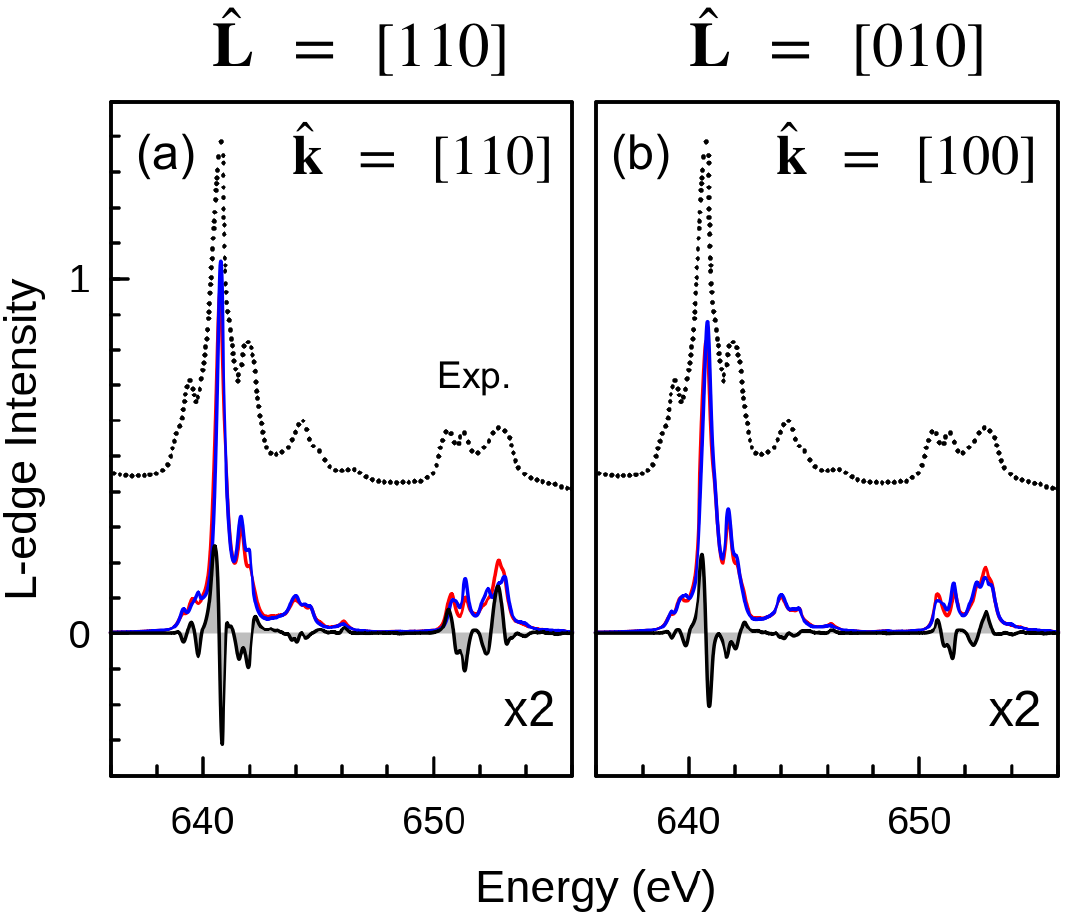}
\caption{The XAS calculated for the two circular polarizations (red and blue) at the Mn $L_{2,3}$ edge together with the XMCD intensities (shaded) for different orientations of the N\'eel vector $\bL$ and x-ray propagation vector $\bf k$. The calculated spectral intensities are broadened by Lorentzian of 0.15~eV (HWHM). The experimental Mn $L_{2,3}$-edge XAS spectrum taken from Ref.~\onlinecite{Groot90} is shown for comparison (the experimental baseline was offset for the sake of clarity.)}
\label{fig_mcd}
\end{figure}

\begin{figure}
\includegraphics[width=0.7\columnwidth]{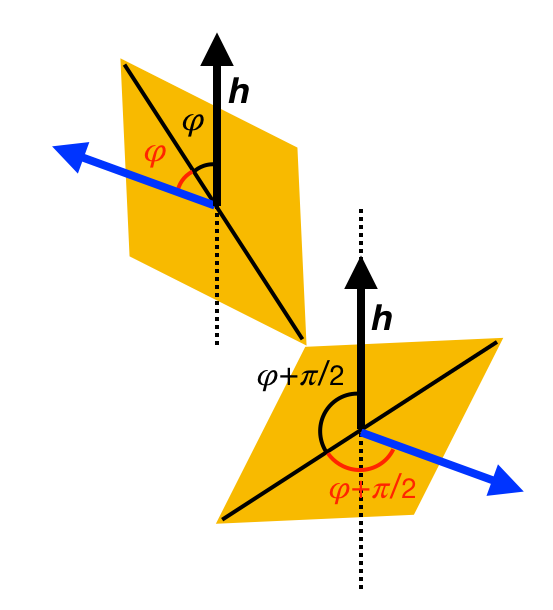}
\caption{
The orientation of the Hall vector $\bh$ (black arrow) and the local spin moments (blue arrows) in the $ab$-plane of the rutile structure.}
\label{fig:cartoon}
\end{figure}

\begin{figure}
\includegraphics[width=1.00\columnwidth]{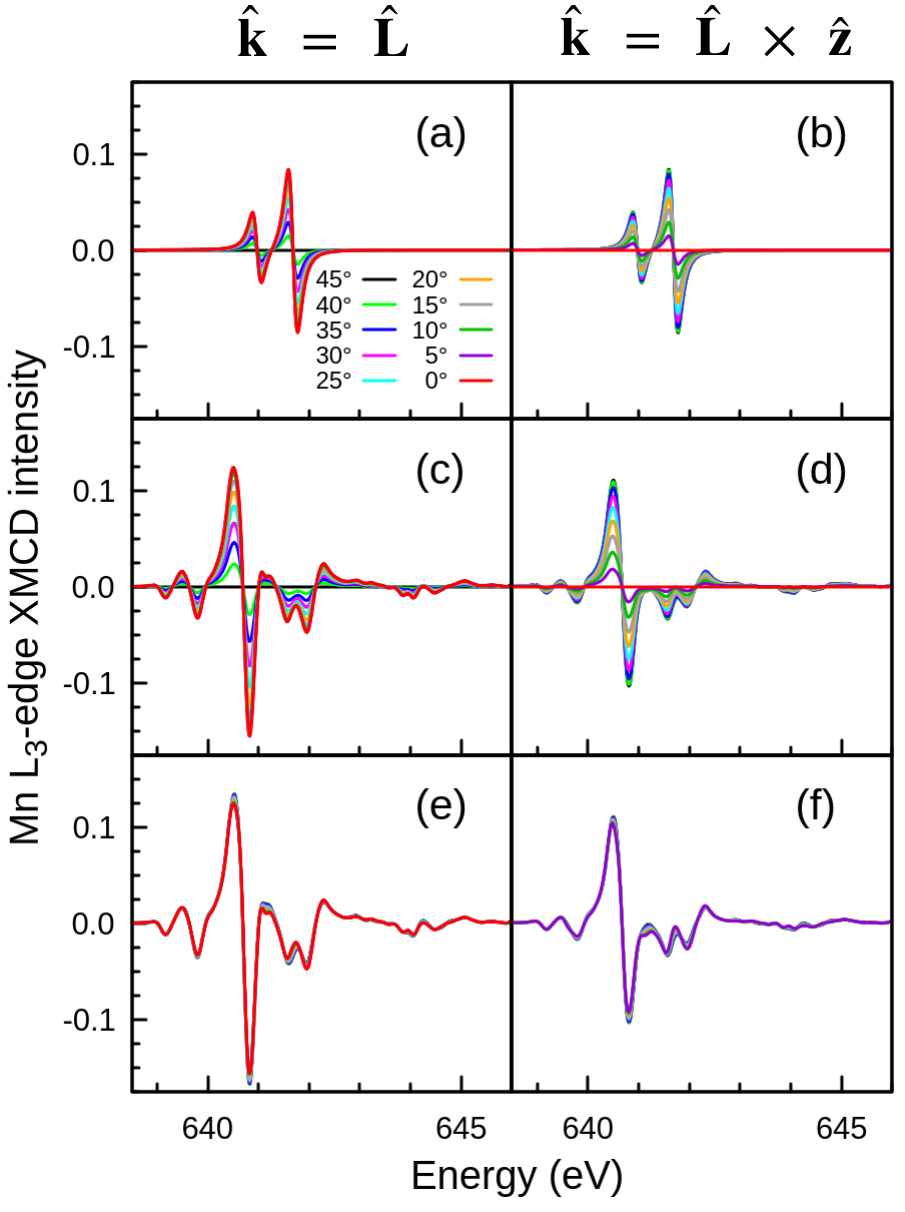}
\caption{The Mn $L_3$-edge XMCD intensities calculated for various angles $\varphi$, with 0$^\circ$ and 45$^\circ$ corresponding to $\hbL = [110]$ and $[010]$, respectively. The valence SOC and CV interaction 
is switched off in (a,b). The XMCD intensities in panels (a,b) collapse onto a single
curve upon division by $\cos(2\varphi)$ and $sin(2\varphi)$, respectively, as required by (\ref{eq:cos}).
In (c,d) the same spectra as in (a,b) calculated with full Hamiltonian are shown. Panels (e,f) show the spectra from (c,d) divided by
$\cos 2\varphi$ and $\sin 2\varphi$, respectively.
}
\label{fig_ldep}
\end{figure}

{\it Results.} Symmetry of the rutile structure~\cite{Smejkal20,Hariki2024a} implies that XMCD is not allowed for $\bL \parallel c$, i.e., $\bh=0$. However, it is allowed if $\bL$ has a finite projection in the $ab$-plane. First, we consider an idealized situation with magnetic moments entirely in the $ab$-plane, i.e., $\bL\perp c$ and $\bbm_1+\bbm_2 = 0$. Figure~\ref{fig_mcd} shows the XAS and XMCD spectra at the Mn $L_{2,3}$ edges calculated for the N\'eel vectors $\hbL= \na$ and $[010]$.
The corresponding Hall vectors are parallel 
($\hat{\bh}= [110]$) and perpendicular ($\hat{\bh}=[100]$) to $\bL$, respectively~\cite{Hariki2024a}. The spectra for $\hat{\bh}= [001]$ a can be found in SM~\cite{sm}.
The XAS line shape agrees well with the experimental data~\cite{Groot90}. Relatively large XMCD intensities are predicted at both the Mn $L_2$ and $L_3$ edges in the present theory, Figs.~\ref{fig_mcd}ab. 

The similarity of the spectra in Figs.~\ref{fig_mcd}ab is not accidental. As shown in Ref.~\onlinecite{Hariki2024a} the two spectra are related by symmetry if the valence SOC and the multipole part of the core-valence (CV) interaction are neglected. This raises questions about the general orientation of $\bL$ in the $ab$-plane. 
In the following we extend the analysis of Ref.~\onlinecite{Hariki2024a} and show that there is a unique relationship between
the in-plane orientation of the N\'eel vector $\bL$ and the XMCD spectrum characterized by the Hall vector $\bh_\bL(\omega)$.
By reversing this relationship, the measured $\bh_\bL(\omega)$ can be used to determine an unknown orientation of $\bL$ in the sample.


In the absence of the valence SOC and the CV interaction, 
the valence spin is decoupled from the rest of the system. 
The states (including excited states) corresponding to 
different orientations of $\bL$ are related by a valence spin rotation. Following Ref.~\onlinecite{Hariki2024a} we fix the coordinate system so that $\hat{\mathbf{x}} = \hbk$ and $\hat{\mathbf{z}}=[001]$, see SM~\cite{sm}, and use dipole operators in the helicity basis along $\hat{\mathbf{z}}$, which is also the quantization axis for spin. 
Eq.~\ref{eq:fsum} for this coordinates becomes
~\footnote{Below, we do not show the frequency dependence and the summation over $f$ explicitly for sake of readability.}
\begin{equation}
    \label{eq:A}
    \begin{split}
       \Delta F(\varphi,\alpha)=&\sum_{f} \expval{f_{\varphi,\alpha}|\hT^{+}-\hT^{-}|i_{\varphi,\alpha}}\expval{i_{\varphi,\alpha}|\hT^0|f_{\varphi,\alpha}}\\
       &\times
       \delta\left(\omega\!-\!E_{fi}\right)+c.c.   \\
       \equiv & (T^+-T^-)\overline{T^0}+c.c.,
           \end{split}
\end{equation}
where the angles $\varphi$ and $\alpha$ capture the orientation of the MnF$_6$ octahedra and the local moment, respectively, relative to the light propagation vector $\hat{\bk}$. The dependence of the single site XMCD on $\varphi$ and $\alpha$ has a simple form~\cite{Hariki2024a,sm}
\begin{equation}
    \begin{split}
    \Delta F(\varphi,\alpha)=&
    (e^{i\alpha}T_{\uparrow}^+
    -e^{-i(2\varphi-\alpha)}T_{\uparrow}^-)
    \overline{T_{\downarrow}^0}\\
    &+
    (e^{i(2\varphi-\alpha)}T_{\downarrow}^+
    -e^{-i\alpha}T_{\downarrow}^-)
    \overline{T_{\uparrow}^0}+
    c.c.
    \end{split}
\end{equation}
Upon summation over the two Mn sites we get
\begin{equation}
\label{eq:cos}
    \begin{split}
    &\tfrac{1}{2}\left(\Delta F(\varphi,\alpha)+\Delta F(\varphi+\tfrac{\pi}{2},\alpha+\pi)\right)\\
    &=e^{i(2\varphi-\alpha)}T_{\downarrow}^+\overline{T_{\uparrow}^0}
    -e^{-i(2\varphi-\alpha)}T_{\uparrow}^-\overline{T_{\downarrow}^0}+c.c.\\
    &=A(\omega)\cos(2\varphi-\alpha),
    \end{split}
\end{equation}
where the bottom line comes from the fact that (\ref{eq:cos}) must vanish for $\varphi=0$ and $\alpha=\tfrac{\pi}{2}$. 
In Fig.~\ref{fig_ldep} we check the validity of (\ref{eq:cos}) by an explicit numerical calculation.

Eq.~\ref{eq:cos} has a simple geometrical interpretation. If one starts with $\bL$ pointing
along $[110]$ or $[1\bar{1}0]$ and rotates it along the $c$-axis, the corresponding Hall vector $\bh(\omega)$ rotates by the same angle in the opposite direction, see Fig.~\ref{fig:cartoon}b, while its $\omega$-dependence remains unchanged, i.e.,
$\hat{\bh}$ is a mirror image of $\hbL$ with respect to $(110)$ or $(1\bar{1}0)$ plane~\footnote{The choice
of $(110)$ vs $(1\bar{1}0)$ plane fixes the sign of the effect and depends on the assignment of
$\bbm_1$ and $\bbm_2$ to the lattice sites.}
\begin{equation}
\label{eq:hL}
    \bh_\bL(\omega)=A(\omega)\hat{\bh} = A(\omega)\mathcal{M}_{[110]}\hbL.
\end{equation}
Note the geometrical meaning of $\hat{\bh}$ as the direction of light propagation for which the XMCD signal is maximal.
This formula applies to any structure with the magnetic sublattices related by a four-fold rotation axis and
another rotation axis perpendicular to it, which determines the zero of $\varphi$ in (\ref{eq:cos}). Without the latter condition an
additional $B(\omega)\sin (2\varphi -\alpha)$ contribution to (\ref{eq:cos}) may appear.

Next, we assess the validity of Eq.~\ref{eq:cos} in the presence of both the valence SOC and the CV interaction. To this end we vary the orientation of $\bL$ within the $ab$-plane and compute the XMCD spectra for
$\hbk=\hbL$ and $\hbk= \hbL \times \hat{\mathbf{z}}$. This corresponds to varying $\varphi$ while fixing $\alpha=0$ and
$\alpha=90^\circ$, respectively, in Eq.~\ref{eq:cos}. Without the valence SOC and the CV interaction the numerical XMCD
spectra perfectly follow $\cos 2\varphi$ and $\sin 2\varphi$ dependencies given by (\ref{eq:cos}), see Fig.~\ref{fig_ldep}ab. With the valence SOC and the CV interaction turned on, the  XMCD spectra in Figs.~\ref{fig_ldep}cd still follow the $\cos 2\varphi$ and $\sin 2\varphi$ dependencies to a good accuracy. This is demonstrated in Figs.~\ref{fig_ldep}ef by dividing the spectra with $\cos 2\varphi$ and $\sin 2\varphi$, respectively, which leads to a collapse on almost identical curves. 
It is not clear to us why the rescaled spectra group into two groups, i.e., why the differences between the curves within the panels e and f are smaller than the difference between the panels. 
We can conclude that, although not exact, Eq.~\ref{eq:hL} is rather fulfilled even for the full Hamiltonian including the valence SOC and the CV interaction.

\begin{figure}[t]
\includegraphics[width=0.98\columnwidth]{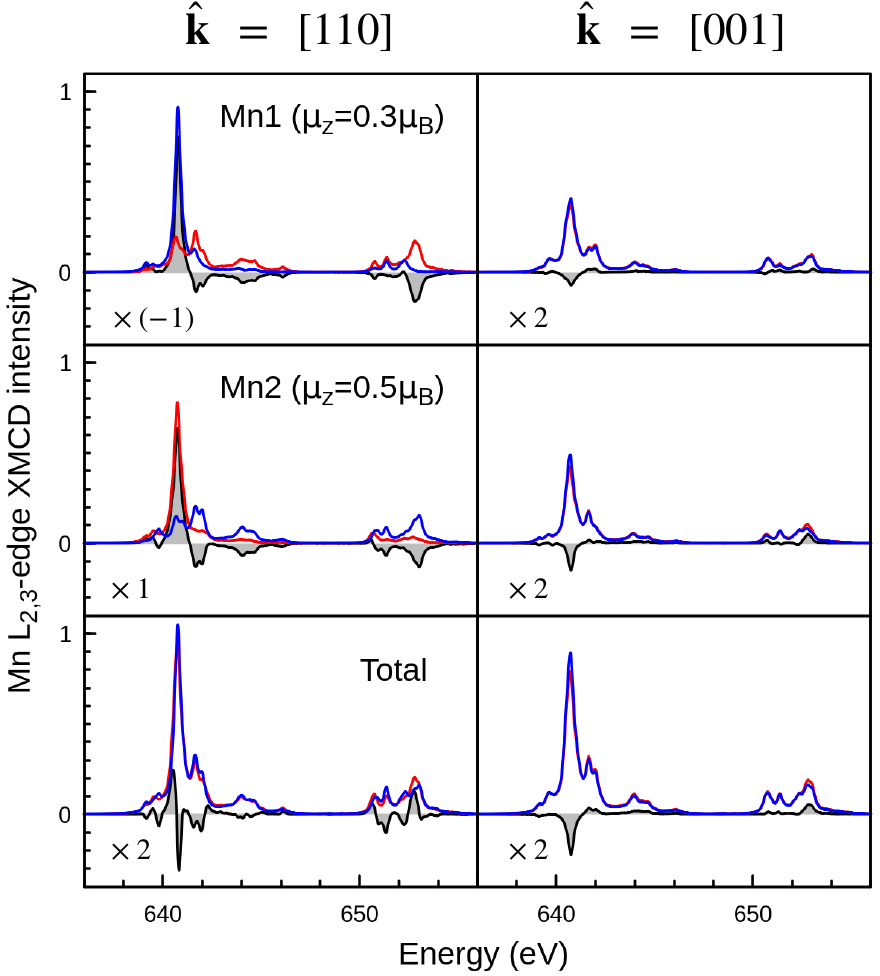}
\caption{The XMCD intensities at the Mn $L_{2,3}$ edge for (left) x-ray propagation vector $\hbk = [110]$ and (right) $\hbk = [001]$ for the Ne\'el vector $\hbL = [110]$ with a small tilt of the Mn magnetic moment along the $z$-axis of $\mu_z=0.5~\mu_B$ at Mn1 (top) and $\mu_z=0.5~\mu_B$ at Mn2 (middle). The total XMCD intensities are shown in the bottom panels.}
\label{fig_tilt}
\end{figure}

Although the easy axis of \mf is parallel to \nc, the N\'eel vector $\bL$ can be flopped into the $ab$-plane by a field of 9--10~T~\cite{King1979,Felcher96} along the $c$-axis, which causes a small canting of the Mn moments $\bS_1$ and $\bS_2$ into the \nc direction. The net magnetization along the $c$-axis for fields close to the spin-flop transition was estimated to $\mu_z=0.3~\mu_{B}$ and 0.5~$\mu_{B}$ in the two Mn sites~\cite{Felcher96}. In Fig.~\ref{fig_tilt}(a), we simulate the effect of canting on the XMCD spectra for $\hbL=\na$. The net magnetization along the $c$-axis gives rise to a finite $h_z$ component of the Hall vector.
The magnitude of XMCD for $\hbk= \nc$ in Fig.~\ref{fig_tilt}(b) is comparable
to the purely altermagnetic effect for $\hbk = \na$ despite the out-of-plane component of the 
local magnetic moments being order of magnitude smaller than in-plane one.
Similar to the experimental observation on MnTe~\cite{Hariki2024b} the altermagnetic ($\hbk = \na$)
and ferromagnetic ($\hbk = \nc$) components of the XMCD spectra exhibit distinct shapes.
This result can be used to estimate the impact of potential misalignment in an experimental set-up.

\begin{figure}
\includegraphics[width=0.98\columnwidth]{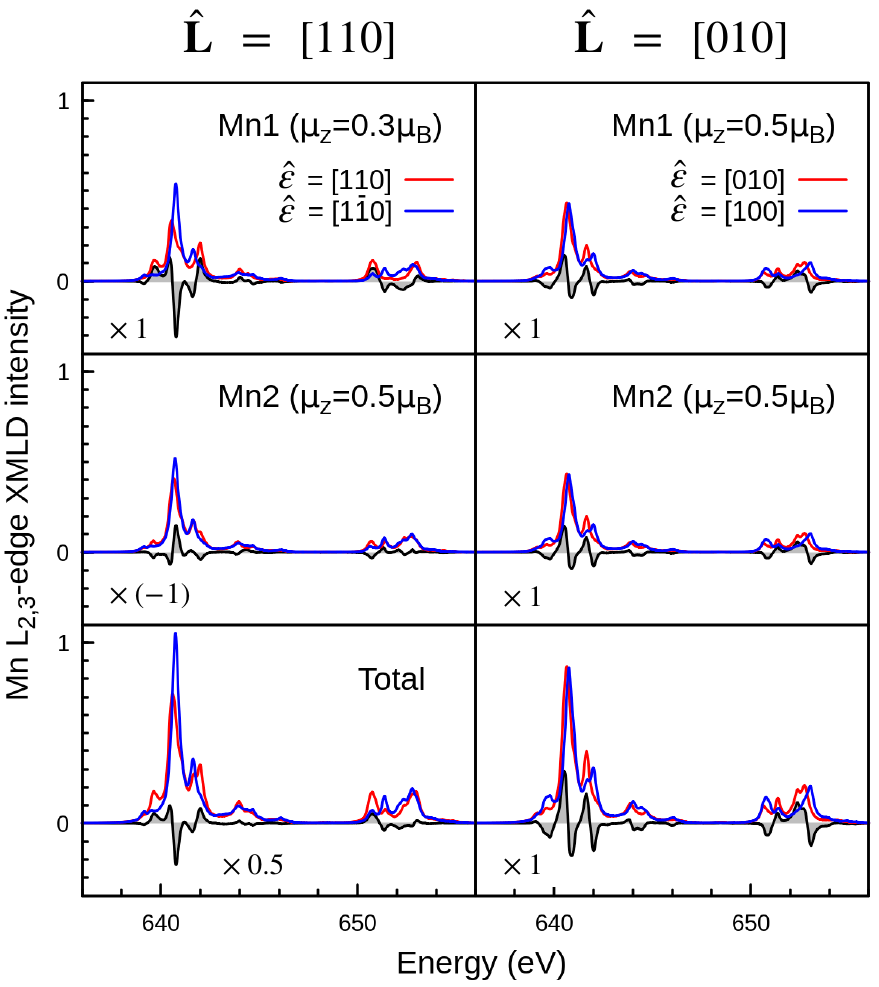}
\caption{The XMLD intensities at the Mn $L_{2,3}$ edge for the x-ray propagation vector $\hbk=\nc$ and the N\'eel vector   $\hbL = \na$ (left) and $\hbL = [010]$ (right). The XMLD is defined for the two x-ray polarization vectors $\boldsymbol{\epsilon}~||~\na$ (red) and $\boldsymbol{\epsilon}~||~[1\overline{1}0]$ (blue). A small tilt of the Mn magnetic moment along the $z$-axis with the values ($\mu_z$) indicated in the panels is considered in the simulation. The total XMLD intensities are shown in the bottom panels.}
\label{fig_xmld}
\end{figure}

Finally, we calculate the linear dichroism (XMLD) in Fig.~\ref{fig_xmld}, which provides the standard x-ray spectroscopic tool to determine the direction of the N\'eel vector in antiferromagnets~\cite{Kuiper93,Saidl2017}, but which cannot distinguish the N\'eel vectors with opposite orientation, which are of particular interest in altermagents.
Our aim is to show the distinct profiles of the XMLD and XMCD spectra, which facilitates the identification and removal of any potential signal contamination arising from imperfect polarization in an experimental implementation.

{\it Discussion.} Next, we compare XMCD in the spin-flopped phase of MnF$_2$ and \amt. Both compounds are $S=5/2$ altermagnetic Mn$^{2+}$ insulators and the presence or absence  of the effect as well as the orientation of the XMCD Hall vector $\bh$) 
depends on the orientation of the N\'eel vector $\bL$, but behaves differently when $\bL$ is rotated in the $ab$-plane. In \amt, $\bh$ points along the $c$-axis and changes sign, vanishes for all frequencies,~\cite{Gonzalez2023,Hariki2024b} at six nodal points. In MnF$_2$, the shape of XMCD spectrum remains approximately unchanged, but $\bh$ rotates in the $ab$-plane in the opposite sense to the rotation of $\bL$.
In both compounds the valence SOC has a minor impact on the XMCD spectra. 
On the other hand, the role of CV multipole interaction is very different due to the different symmetries of the crystal fields in the two compounds. 

In \amt, with $\bh\perp\bL$ geometry,
XMCD completely vanishes if the valence SOC and the CV interaction are absent.
This is caused by the presence of local (3-fold) rotation axis parallel to $\bh$ as explained in Ref.~\onlinecite{Hariki2024b}. 
The $\bh\perp\bL$ geometry takes place also in MnF$_2$
for $\bL\parallel[100]$, however, there is no local rotation axis parallel to 
$\bh$ and there XMCD is allowed even if the valence SOC and the CV interaction are absent. These terms modify the shape 
of the spectra, Fig.~\ref{fig_ldep}, but do not change the magnitude of XMCD substantially.
Therefore the key interaction for the appearance of XMCD in \amt is a mere perturbation
in \mf.
Different origin of the XMCD is also reflected in the XMCD magnitudes, which  
in MnF$_2$ is about five times larger than in \amt.

We have studied the x-ray magnetic circular dichroism in the spin-flop phase of MnF$_2$.
Using an approximate symmetry, we have found a simple relationship between the light propagation vector maximizing the XMCD, $\hat{\bh}$, and the in-plane N\'eel vector $\bL$ in the rutile structure, which allows a unique
determination of $\bL$ from the angular dependence of XMCD.
Comparing \mf and \amt, we have shown that even in isoelectronic compounds XMCD may originate in different terms in the Hamiltonian
depending on their symmetries.

\begin{acknowledgements}
We thank Karel V\'yborn\'y, Andriy Smolyanyuk and Anna Kauch for discussions and critical reading of the manuscript.
This work was supported by JSPS KAKENHI Grant Numbers 21K13884, 21H01003, 23K03324, 23H03817 (A.H.), and by the project Quantum materials for applications in sustainable technologies (QM4ST), funded as project No. CZ.02.01.01/00/22\_008/0004572 by Programme Johannes Amos Commenius, call Excellent Research and
by the Ministry of Education, Youth and Sports
of the Czech Republic through the e-INFRA CZ (ID:90254)(J.K.).
\end{acknowledgements}

\bibliography{main}

\end{document}


\title{Supplementary Material to ``Determination of the N\'eel vector in rutile altermagnets through x-ray magnetic circular dichroism: the case of \mf''}

\author{A.~Hariki}
\affiliation{Department of Physics and Electronics, Graduate School of Engineering,
Osaka Metropolitan University, 1-1 Gakuen-cho, Nakaku, Sakai, Osaka 599-8531, Japan}
\author{T.~Okauchi}
\affiliation{Department of Physics and Electronics, Graduate School of Engineering,
Osaka Metropolitan University, 1-1 Gakuen-cho, Nakaku, Sakai, Osaka 599-8531, Japan}
\author{Y.~Takahashi}
\affiliation{Department of Physics and Electronics, Graduate School of Engineering,
Osaka Metropolitan University, 1-1 Gakuen-cho, Nakaku, Sakai, Osaka 599-8531, Japan}
\author{J.~Kune\v{s}}
\affiliation{Department of Condensed Matter Physics, Faculty of
  Science, Masaryk University, Kotl\'a\v{r}sk\'a 2, 611 37 Brno,
  Czechia}

\maketitle

\section{XMCD simulation}

\begin{figure}[b]
\includegraphics[width=0.50\columnwidth]{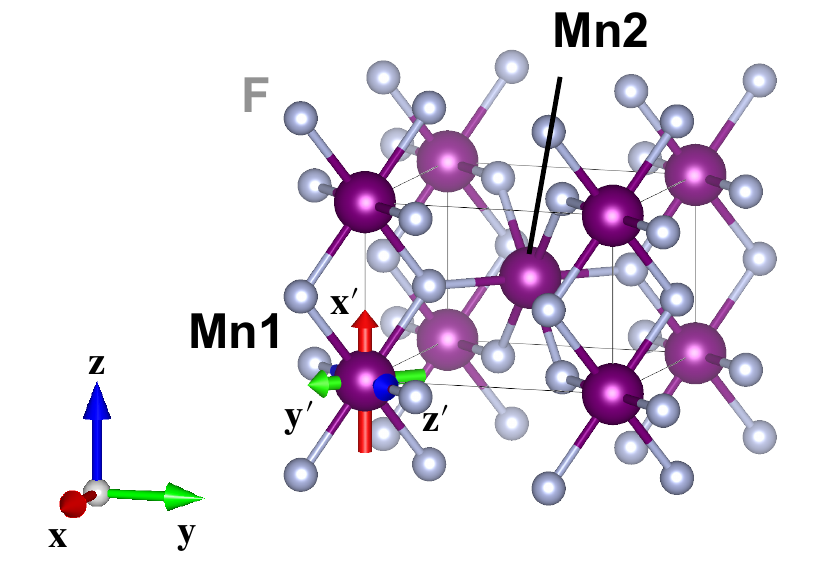}
    \caption{The crystal structure of MnF$_2$ and a local coordinate (x'y'z').}
\label{fig_struct_local}
\end{figure}

\begin{figure}[b]
\includegraphics[width=0.97\columnwidth]{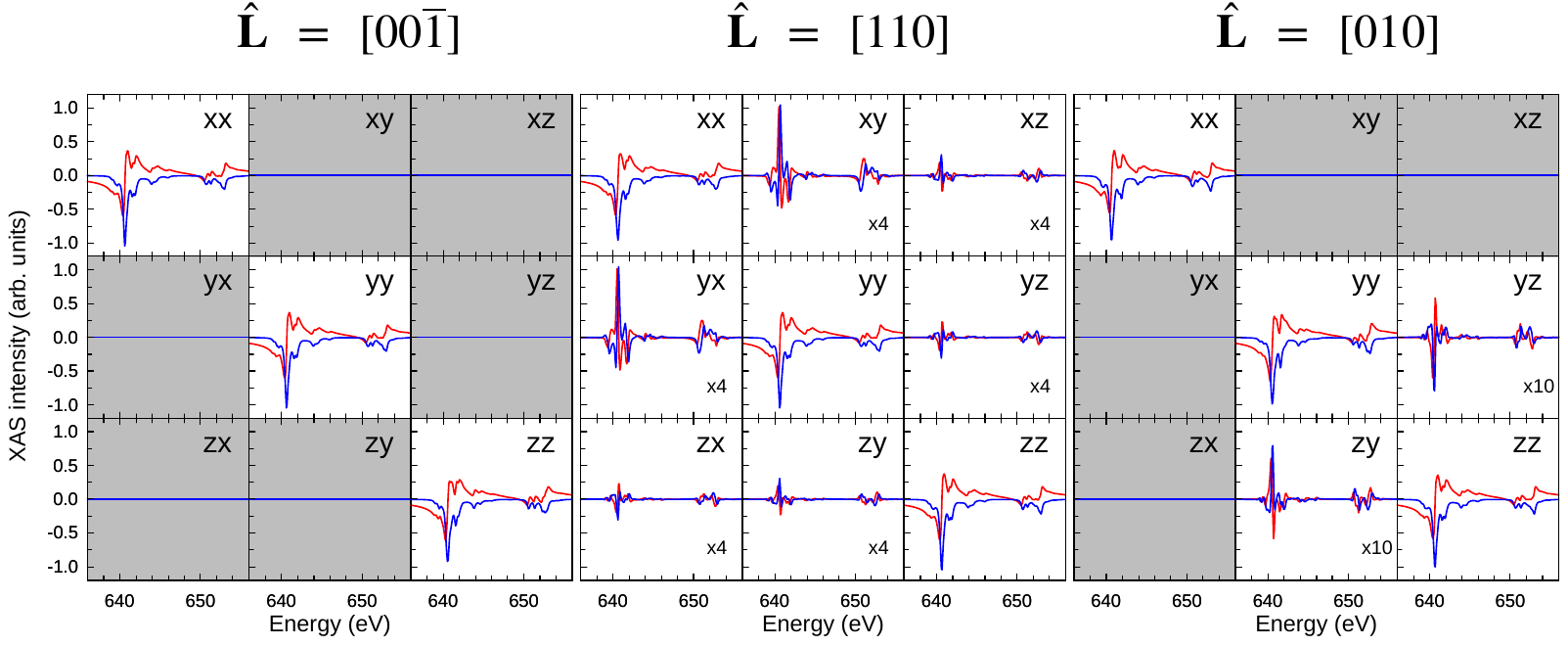}
    \caption{Real (red) and imaginary (blue) part of the optical conductivity tensor at Mn $L_{2,3}$ edge for different orientations of the N\'eel vector $\bL$.}
\label{fig_abc}
\end{figure}

We calculate Mn $L$-edge x-ray magnetic circular dichroism (XMCD) and x-ray magnetic linear dichroism (XMLD) spectra using the Mn$^{2+}$ atomic model. First, we perform the density functional theory (DFT) calculation using the WIEN2K packages~\cite{wien2k} with the local density approximation for the exchange correlation potential. 
The experimental rutile structure was adopted in the DFT calculation.
The crystal-field Hamiltonian $h^{\text{CF}}$ in the atomic model is derived from the DFT bands by projecting the Mn 3$d$ bands near the Fermi level onto a tight-binding model using the wannier90 and wien2wannier packages~\cite{wannier90,wien2wannier}, as described in Ref.~\cite{Hariki2024a}.
The $h^{\text{CF}}$ is represented in a crystal-field basis ($x'y'$, $3z'^2-r'^2$, $x'^2-y'^2$, $z'x'$, $y'z'$) for a local coordinate ($x'y'z'$) in Fig.~S\ref{fig_struct_local} as,
\begin{equation*}
h^{\text{CF}}=
\begin{pmatrix}
0.615 & 0 & 0 & 0 & 0 \\
0 & 0.550 & 0.097 & 0 & 0 \\
0 & 0.097 & -0.114 & 0 & 0 \\
0 & 0 & 0 & -0.142 & 0 \\
0 & 0 & 0 & 0 & -0.120 \\
\end{pmatrix}.
\end{equation*}
The core-valence interaction and SOC parameters are estimated by an atomic Hartree-Fock calculation following Refs.~\cite{Hariki2017,Hariki20}.
The Slater integrals for the multipole part in the $dd$ (valence-valence) and $pd$ (core-valence) interaction in the present Mn$^{2+}$ atomic model are summarized in Table~\ref{tab_slater}~\footnote{In the atomic model the monopole parts $F_0^{dd}$ and $F_0^{pd}$ contribute only an overall shift to the spectra and thus are irrelevant in our treatment.}.
The XAS spectral function is described by the Fermi's golden rule 
\begin{equation}
\label{eq:xas}
\begin{split}
F^{(i)}_{\rm XAS}(\omega_{\rm in})&=-\frac{1}{\pi}\operatorname{Im} \langle g | T^{\dag} \frac{1}{\omega_{\rm in}+E_g-H_i} T| g \rangle. \notag \\
\end{split}
\end{equation}
Here, $|g\rangle$ is the ground with energy $E_g$, and $\omega_{\rm in}$ is the energy of the incident photon. 
$T$ is the electric dipole transition operator representing circularly and linearly-polarized x-rays for XMCD and XMLD, respectively~\cite{Matsubara00}. The index $i$ ($=1,2$) denotes the x-ray excited Mn site in the unit cell. The atomic Hamiltonian $H_{i}$ for each Mn site is used to compute the individual site contribution, and the total XMCD intensities are obtained by summing up the contributions of two sites. To complement the main text, in Fig.~\ref{fig_abc} we show the elements of the conductivity tensor $\boldsymbol{\sigma}(\omega)$ at the Mn $L$-edge for the N\'eel vector along $[00\overline{1}]$, [110], and $[010]$ directions.

\begin{figure}[t]
\includegraphics[width=0.50\columnwidth]{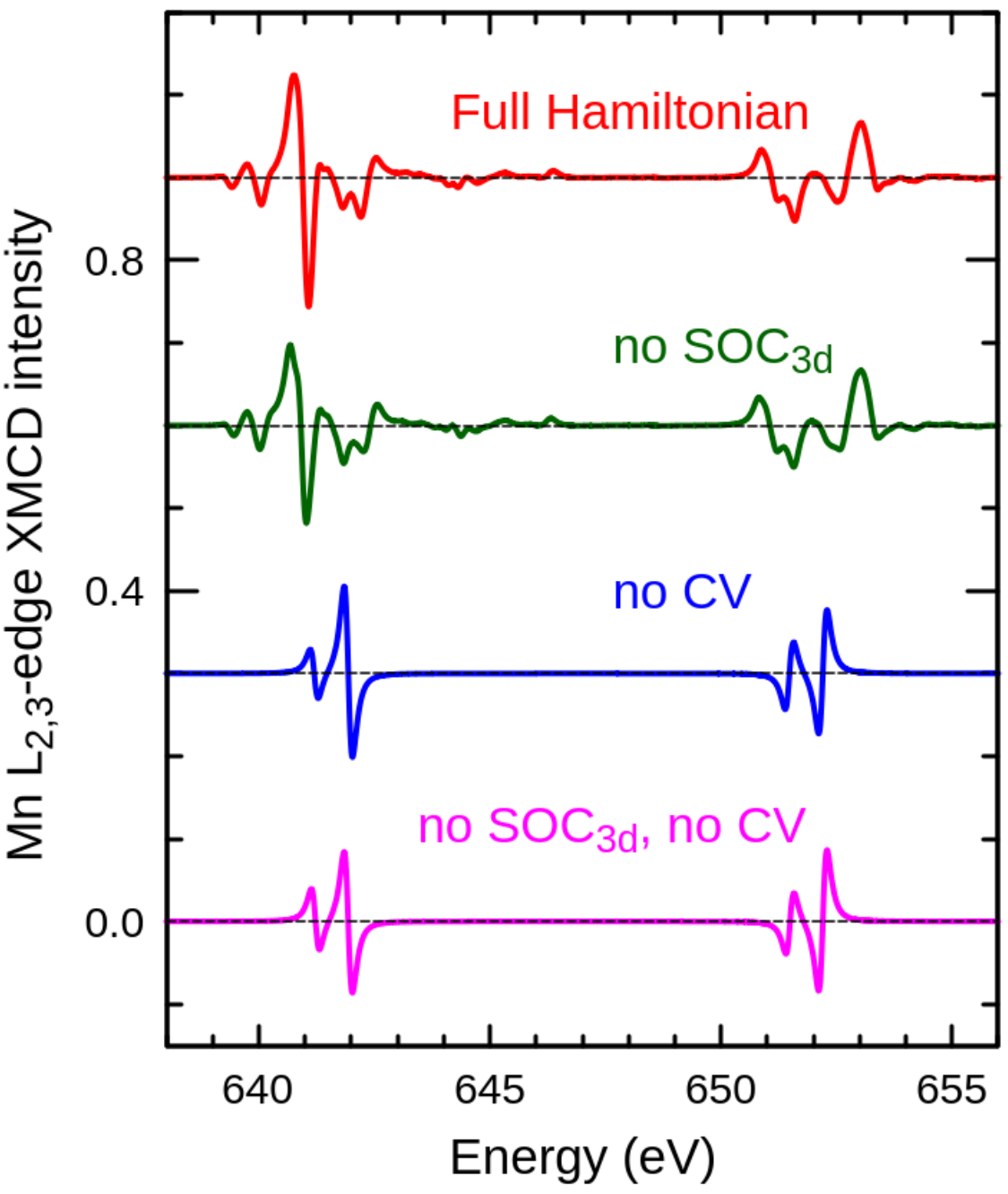}
\caption{The XMCD intensities at the Mn $L_{2,3}$ edge for $\bk, \bL \parallel [110]$ for the models with (red) full Hamiltonian, (green) no 3$d$ SOC, (blue) no 2$p$--3$d$ core-valence (CV) multiplet interaction, and (pink) no 3$d$ SOC and CV.}
\label{fig_mcd_nsoc}
\end{figure}

In Fig.~\ref{fig_mcd_nsoc}, we calculated the Mn $L$-edge XMCD with a modified Mn$^{2+}$ atomic model Hamiltonian which neglects the Mn valence SOC and/or the core-valence (CV) interaction terms.  The valence SOC has almost no effect on the XMCD intensities and shapes. The core-valence interaction is important for the XMCD spectral shape, but large XMCD intensities remain without the CV term.

      \begin{table}[h]
            \centering  
            \begin{tabular}{c | c c c c c c }
            \hline\hline\\[-10pt]
            &{\ }
            $F^{dd}_{2}${\ }  \, & {\ }$F^{dd}_{4}${\ } \, & {\ }$F^{pd}_{2}${\ } \, & {\ }$G^{pd}_{1}${\ } \, & {\ }$G^{pd}_{3}${\ }
            \\[+4pt]
            \hline\\[-7pt]
            {\quad}Values (eV){\ }{\quad}&{\ }7.409{\ }&{\ }4.631{\ }&{\ }6.321{\ }&{\ }4.606& {\ }2.618{\ }\\[+7pt]
            \hline\hline
            \end{tabular}
            \vspace*{+0.2cm}
            \caption{The atomic Slater integral values for the $d$-$d$ (valence-valence) and $p$-$d$ (core-valence) interaction employed in the Mn$^{2+}$ atomic model. Here, $F$ and $G$ denote the direct and exchange integrals, respectively.}
            \label{tab_slater}
        \end{table}        

\section{XMCD with no core-valence exchange and no valence SOC}
\label{app:nsoc}
For the sake of completeness we summarize the derivations of Ref.~\onlinecite{Hariki2024a} below.
We start with several definitions. The circular dichroism ${\Delta F(\hat{\bk},\bS)=
F^+(\hat{\bk},\bS)-F^-(\hat{\bk},\bS)}$ is the difference of the absorption spectra for the 
right-hand and left-hand circularly polarized light can be obtained from 
the Fermi golden rule
\begin{equation}
\label{eq:fsum}
F^\pm(\hat{\bk},\bS)=\sum_{f} \left|\expval{f_\bS|\hT^{\pm}_{\hat{\bk}}|i_\bS}\right|^2\!\delta\left(\omega\!-\!E_{fi;\bS}\right).
\end{equation}
Here $|i_\bS\rangle$ and $|f_\bS\rangle$ are the eigenstates of the Hamiltonian, $E_{fi;\bS}$ is the excitation energy, and $\hT^{\pm}_{\hat{\bk}}$ are the dipole operators for the right- and left-hand polarization
with respect to propagation vector $\hat{\bk}$.
Thanks to the immobility of the core hole
the x-ray absorption spectrum is an sum over site contributions.

\begin{figure}
    \centering
    \includegraphics[width=0.5\linewidth]{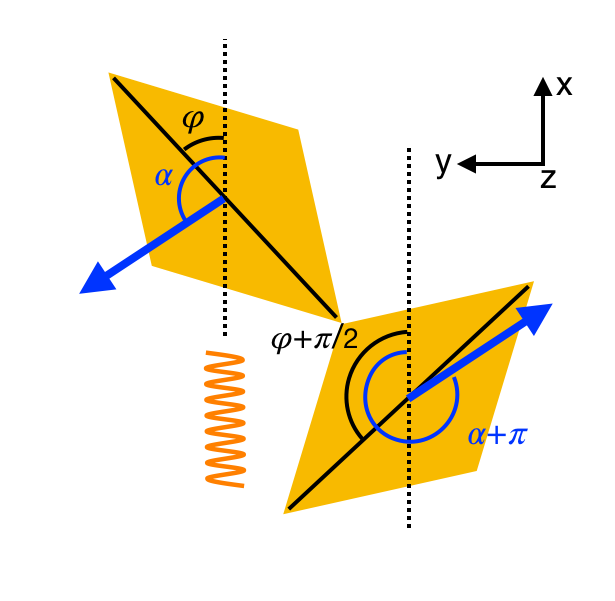}
    \caption{Geometry used to study the in-plane $\bL$ and $\bk$ configurations for XMCD. The blue arrows indicate the local spin moments $\bS_{1,2}$.
    Note that the coordinate system used here is chosen so that $\hat{\mathbf{z}}=\hat{\mathbf{c}}$ and $\hat{\mathbf{x}}=\hat{\bk}$.}
    \label{fig:Scartoon}
\end{figure}

We will use the geometry of Fig.~\ref{fig:Scartoon} with the x-rays coming along the $x$-axis
and the quantization $z$-axis of spin and angular momenta parallel to the
crystallographic $c$-axes. To evaluate XMCD for 
 various  orientation of the spin moment $\bS$ and incoming light direction $\bk$ we will
vary the angles $\varphi$ (orientation of the crystal) and $\alpha$ (orientation of the local spin moment).
Using the relations between the dipole operators $\hT^x$, $\hT^y$ and $\hT^z$ 
    \begin{equation*}
    \begin{split}
        T^+&=
        -T^x-iT^y
        \\
        T^-&=
        -T^x+iT^y
        \\
        T^0&=T^z,
    \end{split}
    \end{equation*}
we express the dipole operators $T^\pm_{\hat{x}}$ for the circularly polarizated light propagating along the $x$-axis,
helicity basis for $\hat{\bk}\parallel x$,  using dipole operators 
in the helicity basis for $\hat{\bk}\parallel z$\footnote{We drop the $\hat{z}$ subscript in $T^\pm_{\hat{z}}$ for sake of readability.}
\begin{equation*}
    T^\pm_{\hat{x}}=-T^y\mp iT^z=i\left(\frac{T^+-T^-}{2}\pm T^0\right).
\end{equation*}
The expression (\ref{eq:fsum}) for XMCD then takes the form  
    \begin{equation}
    \label{eq:A}
    \begin{split}
       \Delta F(\varphi,\alpha)=&\sum_{f} \expval{f_{\varphi,\alpha}|\hT^{+}-\hT^{-}|i_{\varphi,\alpha}}\expval{i_{\varphi,\alpha}|\hT^0|f_{\varphi,\alpha}}\\
       &\times
       \delta\left(\omega\!-\!E_{fi}\right)+c.c.   \\
       \equiv & (T^+-T^-)\overline{T^0}+c.c.
           \end{split}
    \end{equation}
    Here, the polarization of the dipole operators is taken with respect to the $z$-axis. We will use
    the shorthand notation for the matrix elements of the dipole operators shown on the third line from now on.
    In absence of the valence SOC  and core-valence multipole interaction, the excitation energy
    does not depend on the orientation of $\bL$, i.e., the angle $\alpha$ and the eigenstates
    for arbitrary angles $\varphi$ and $\alpha$ can be obtained
from those for $\varphi=0$ and $\alpha=0$ by 
a joint $z$-axis rotation $\mathcal{C}(\varphi,\alpha)$ of core spin, core orbitals and valence orbitals by angle $\varphi$ and rotation of
valence spin by angle $\alpha$, which transform the operators
\begin{equation*}
    \mathcal{C}(\varphi,\alpha): d^{\phantom\dagger}_{ms}\rightarrow 
    e^{im\varphi}e^{i\sigma\alpha}d^{\phantom\dagger}_{m\sigma},
    \quad
    p^{\phantom\dagger}_{m\sigma}\rightarrow 
    e^{i(m+\sigma)\varphi}p^{\phantom\dagger}_{m\sigma},
\end{equation*}
where $m$ is the orbital and $\sigma=\pm\tfrac{1}{2}$ the spin projection 
along the $z$-axis. Next, we use
\begin{equation*}
    \expval{\mathcal{C}(\varphi,\alpha)f|\hT|
    \mathcal{C}(\varphi,\alpha)i}=
    \expval{f|\mathcal{C}^{-1}(\varphi,\alpha)\hT
    \mathcal{C}(\varphi,\alpha)|i}
\end{equation*}
to transform the dipole operators instead of wave functions in (\ref{eq:A}). 
Using the definition of the dipole operators
\begin{equation*}
\label{eq:F}
\begin{split}
 \hT^{\pm}&\equiv \hT_\uparrow^{\pm}+ \hT_\downarrow^{\pm}=\sum_{m,\sigma}\Gamma^{\phantom n}_{\pm m} \hd^\dagger_{m\pm 1\sigma}\hp^{\phantom\dagger}_{m\sigma}\\
 \hT^{0}&\equiv\hT_\uparrow^{0}+ \hT_\downarrow^{0}=\sum_{m,\sigma}\Gamma^{(0)}_{m} \hd^\dagger_{m\sigma}\hp^{\phantom\dagger}_{m\sigma},
\end{split}
\end{equation*}
we arrive at their transformation properties
    \begin{equation}
        \begin{split}
     \mathcal{C}^{-1}(\varphi,\alpha)\hT_\sigma^{+}\mathcal{C}(\varphi,\alpha) & =  e^{i\varphi} e^{-i\sigma(\varphi-\alpha)}\hT_\sigma^{+}\\
     \mathcal{C}^{-1}(\varphi,\alpha)\hT_\sigma^{-}\mathcal{C}(\varphi,\alpha) & =  e^{-i\varphi} e^{-i\sigma(\varphi-\alpha)}\hT_\sigma^{-}\\
     \mathcal{C}^{-1}(\varphi,\alpha)\hT_\sigma^{0}\mathcal{C}(\varphi,\alpha) & =  e^{-i\sigma(\varphi-\alpha)}\hT_\sigma^{0}.
     \end{split}
\end{equation}
Substituting these into (\ref{eq:A}) we get the  formula for the XMCD single-site spectra for angles $\varphi$ and $\alpha$
\begin{equation}
    \begin{split}
    \Delta F(\varphi,\alpha)=&(e^{i\varphi}T_{\uparrow}^+
    -e^{-i\varphi}T_{\uparrow}^-)
    \overline{T_{\uparrow}^0}\\
    &+
    (e^{i\varphi}T_{\downarrow}^+
    -e^{-i\varphi}T_{\downarrow}^-)
    \overline{T_{\downarrow}^0}\\
    &+
    (e^{i\alpha}T_{\uparrow}^+
    -e^{-i(2\varphi-\alpha)}T_{\uparrow}^-)
    \overline{T_{\downarrow}^0}\\
    &+
    (e^{i(2\varphi-\alpha)}T_{\downarrow}^+
    -e^{-i\alpha}T_{\downarrow}^-)
    \overline{T_{\uparrow}^0}+
    c.c.\\
    =&
    (e^{i\alpha}T_{\uparrow}^+
    -e^{-i(2\varphi-\alpha)}T_{\uparrow}^-)
    \overline{T_{\downarrow}^0}\\
    &+
    (e^{i(2\varphi-\alpha)}T_{\downarrow}^+
    -e^{-i\alpha}T_{\downarrow}^-)
    \overline{T_{\uparrow}^0}+
    c.c.
    \end{split}
\end{equation}
The $\uparrow\uparrow$ and $\downarrow\downarrow$ terms do not change sign under the magnetic moment
reversal $\alpha \rightarrow \alpha +\pi$ and thus must vanish the expression for XMCD.

\bibliography{main}